\documentclass[11pt,a4paper]{article}
\usepackage{jinstpub}
\usepackage{lipsum}  
\usepackage{subcaption}
\usepackage{here}
\usepackage[dvipsnames]{xcolor}
\usepackage[T1]{fontenc}
\usepackage{tgcursor}
\usepackage{lineno}
\usepackage{verbatim}
\usepackage{graphbox}
\usepackage{float}

\title{MPGD with a 3D-printed Thick-GEM as sole gain element
}

\author[a]{J.~Collins,}
\author[a,1]{M.~Hohlmann\note{Corresponding author.}}

\affiliation[a]{Dept.\ of Aerospace, Physics, and Space Sciences, Florida Institute of Technology, Melbourne, FL 32901, USA}

\emailAdd{hohlmann@fit.edu}

\abstract{

We present the first micropattern gaseous detector that employs a small 3D-printed Thick-GEM as its sole gain element. The detector can achieve sufficient gas gain for regular operation without the need for pre-amplification by additional gain elements. We describe the design, quality control, assembly, and test of this detector. The 10~cm $\times$ 10~cm active area of the Thick-GEM features three separate sectors with different sizes  of the clearance rim annuli (0.1~mm, 0.15~mm, 0.2~mm) around the 0.7~mm diameter holes. The gas gain is found to depend strongly on the rim size. When operated in Ar/CO$_2$ 70:30 gas, the sector with 0.15~mm annulus rims reaches a gain above 10$^4$ while operating in a stable manner with an acceptably low discharge rate. The gain reach under stable operation is found to be considerably lower in the other two sectors. The gas gain shows a characteristic time-dependence as it rises quickly in the first hour of operation and then drops slowly over the next 24 hours and subsequently stabilizes.
}


\begin{document}

\maketitle
\flushbottom
\pagebreak
\section{Introduction}\label{sec:int}

The global procurement sources for micro-pattern gaseous detectors (MPGDs)~\cite{Sauli:2021} are currently rather limited, with the CERN Micro-Pattern Technologies workshop essentially being the sole stable supplier. Repeated efforts by the gaseous detector R\&D community to move MPGD production to industry and to create a lasting commercial source for MPGDs have mostly failed -- presumably due to the lack of a sizable commercial market for these devices.
At the same time, additive manufacturing or “3D-printing” has been actively used for quite some time to produce mechanical support structures for particle detectors in the high energy physics and nuclear physics communities. 
If active MPGD gain elements could also be 3D-printed, then a path would be open to 3D-print entire MPGDs, possibly in one go. This could potentially speed up detector development and production tremendously and save cost. It would also allow end users to produce MPGDs in house by simply printing them out. 

However, there has been only modest progress in using this technology for producing the active gain elements in MPGDs since it was suggested to pose  such a development as a "grand challenge" to the gaseous detector community over a decade ago~\cite{Hohlmann:2013hda}. The main obstacle is the difficulty with simultaneously 3D-printing conductive materials and insulating materials in the same object. Industry began producing printers with that capability at the end of the previous decade with the purpose of 3D-printing circuit boards. The first 3D-printed gain element for an MPGD was created at CERN in 2019~\cite{Brunbauer:2019ubp} in the form of a Thick-GEM~\cite{Breskin:2008cb} structure. Optical detection of scintillation light in the operating gas mixture established the existence of an electron avalanche inside the holes of the Thick-GEM that reached a gas gain of factor five. However, the detector was unstable at higher voltages preventing measurements with higher gain. Consequently, two standard gas electron multipliers (GEMs)~\cite{Sauli:1997qp,Sauli:2016eeu} were used as additional amplification elements to create sufficient gas gain for X-ray spectrum measurements. Poor electron transfer through the holes (perhaps caused by documented printing anomalies) was given as the cause for the inability to record energy spectra that resolve the $^{55}$Fe photo peak and escape peak. 

In this paper, we present the design and performance of the first 3D-printed THGEM that achieves sufficient gain to work as the sole gain element in an MPGD without the need for additional pre-amplification. A particular focus of our studies is the operational stability of the device.

\section{Thick-GEM Detectors}\label{sec:THGEMs}
Thick-GEMs (THGEMs) are a particular type of MPGD, which are significantly thicker versions of GEMs and made of sturdier materials. They are typically manufactured by drilling holes into a standard printed circuit board (PCB). Instead of the double-conical hole shape of standard GEMs, THGEMs use straight holes with clearance rims, i.e.\ regions around the holes free of the metal which otherwise covers the full surface of the active area of the THGEM board. The size of these rims can have significant effects on the gain and on its long-term behavior under irradiation~\cite{Sauli:2021}. The holes of THGEMs are also about an order of magnitude larger than holes in standard GEMs. THGEMs are robust and can be produced significantly cheaper than GEMs while producing comparable gains and achieving decent position resolution, as well as having high-rate capability and fast signals~\cite{Breskin:2008cb}. The mechanical properties of THGEMs, with their thicker supports, make them useful for applications requiring large areas and rigid electrodes, such as photosensitive detectors~\cite{Alexeev:2015kda} and cryogenic devices~\cite{Bondar:2012qz}.

\section{Thick-GEM Design}
\begin{figure}[H]
\centering
\includegraphics[width=0.9\linewidth]{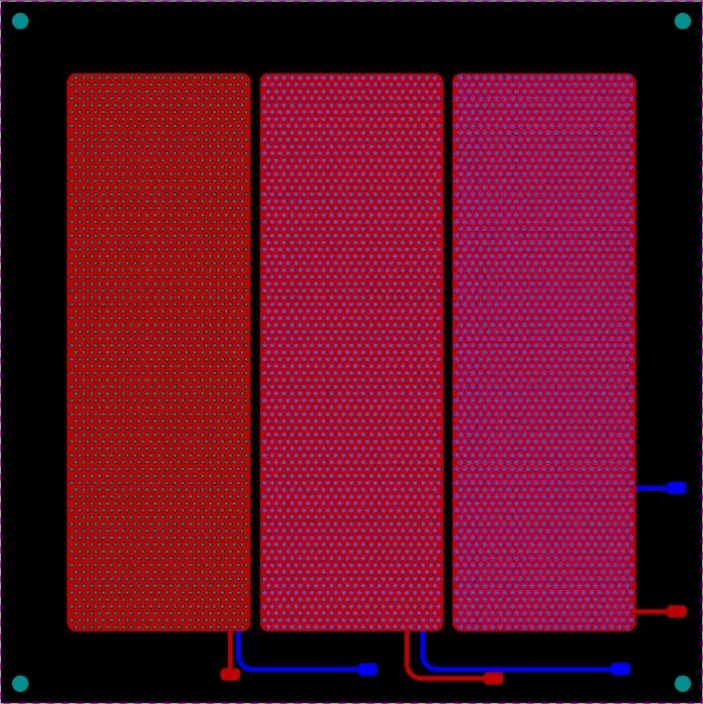}
\caption{\label{THGEM Altium Model}View of the 10~cm $\times$ 10~cm THGEM design within Altium Designer. The front (back) electrodes are show in red (blue).}
\end{figure}

Instead of drilling holes into a standard PCB, we produce a 10~cm $\times$ 10~cm THGEM through 3D printing with a commercial inkjet printer that uses a proprietary formulation for a conductive ink based on silver~\cite{nanodim}. With standard THGEM production methods, the rims around the holes are formed through wet etching. In our case, the clearances around the holes are directly implemented in the design. The board layout is designed using Altium PCB design software~\cite{altium}.
The 10 cm $\times$ 10 cm active region is divided into three equal sectors (Fig.~\ref{THGEM Altium Model}). Each of these sectors has its own traces leading to pads for connection to separate channels in an HV power supply so that each sector can be powered independently, creating three separate types of THGEMs on one board.

\begin{figure}[H]
\centering
\includegraphics[width=0.9\linewidth]{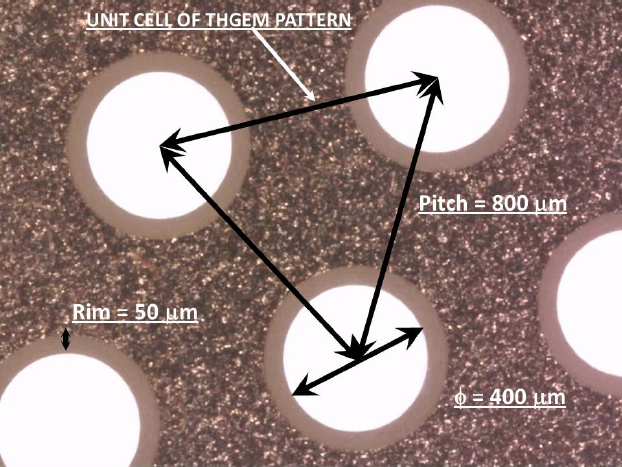}
\caption{\label{Thick GEM Parameters}Typical hole dimensions and pitches for standard THGEMs (reproduced from \cite{Alexeev:2015kda}).}
\end{figure}

The design of the hole dimensions follows various considerations presented in Ref.~\cite{Alexeev:2015kda}. All holes in each sector have the same diameter of 0.7~mm, with a pitch of 1.4~mm. The board thickness is designed to be 0.7~mm as well, which means that the largest hole cross-section is square. These dimensions are on the larger side for THGEMs, as standard THGEMs typically have PCB thicknesses of 0.2-1.0~mm, hole diameters of 0.2-1~mm, pitches of 0.5-1.2~mm, and rim widths between 0 and 0.1~mm \cite{Alexeev:2015kda}.  An example is shown in Fig.~\ref{Thick GEM Parameters}. Our THGEM dimensions are chosen in this way to ease the precision requirements for the 3D-printing process.

Different rim sizes are used in our THGEM design, which constitute the defining difference between the three sectors of the overall board. Previous studies have indicated that the rims provide not only a general function similar to that of the double-conical hole design of standard GEMs, but differences in rim annulus size affect the overall effective gain of the detector, as well as the time evolution of the gain~\cite{Alexeev:2015kda}. Consequently, the rim annuli in the three sector of our board are designed to have three different values; specifically, they are 0~mm (no rim), 0.1~mm, and 0.18~mm.

\section{3D-printing the Thick-GEM}
\begin{figure}[t]
\centering
\includegraphics[height=7cm, width=0.49\linewidth, trim={3cm, 0cm, 3cm, 0.7cm}, clip]{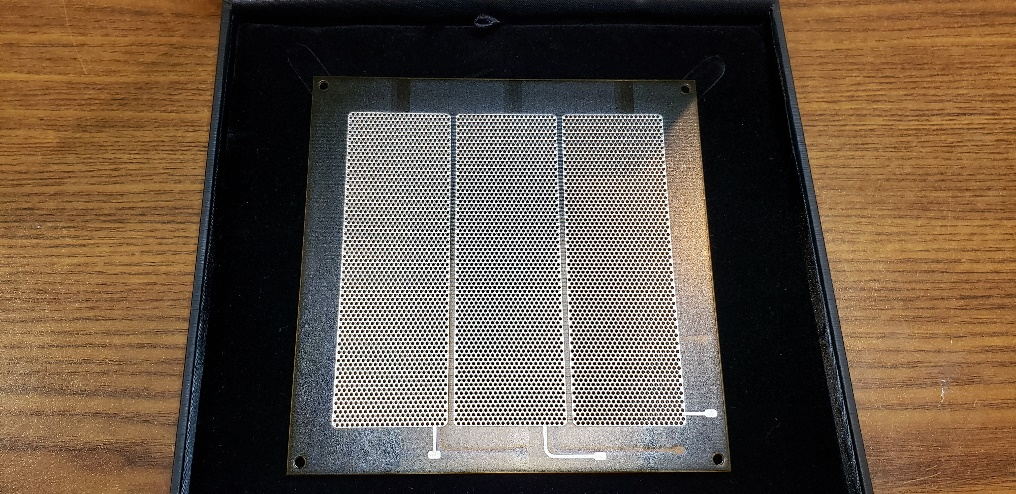}
\includegraphics[height=7cm, width=0.49\linewidth, trim={3.5cm, 0cm, 2cm, 0cm}, clip]{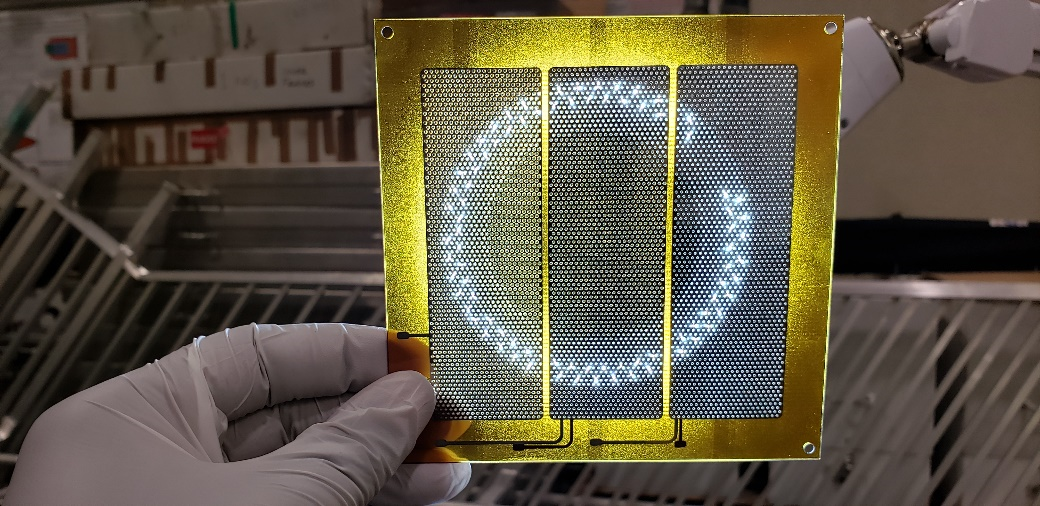}
\caption{\label{THGEM light}Left: 3D-printed THGEM as delivered by printer company. Right: Light from a ring light shining through the THGEM.}
\end{figure}

\begin{figure}
\centering
\includegraphics[width=0.9\linewidth]{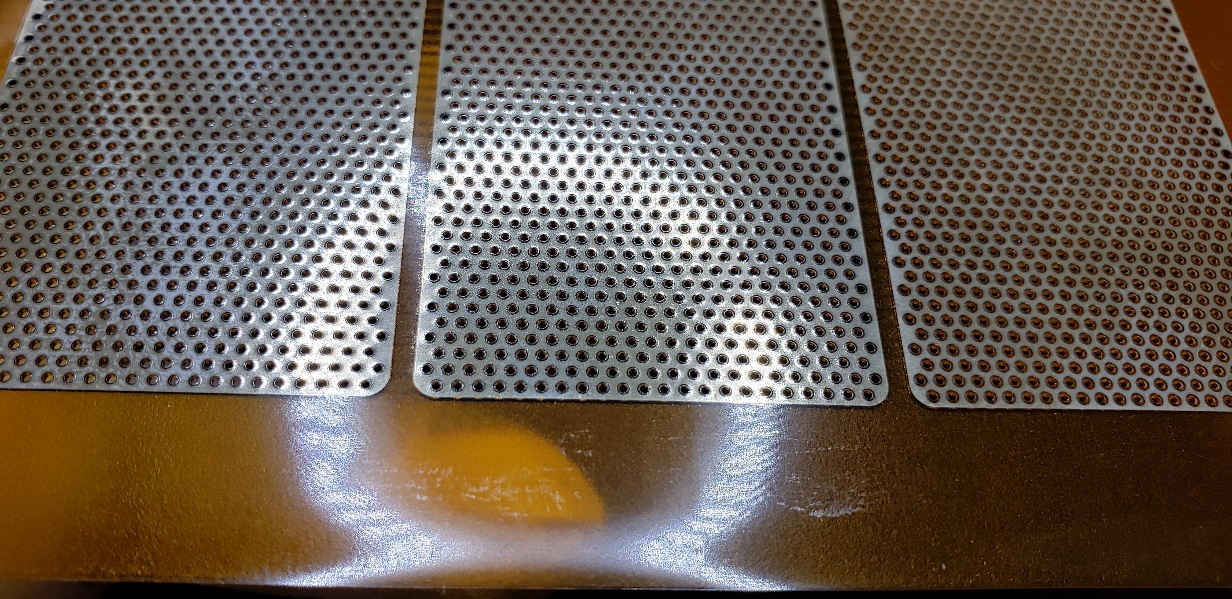}
\caption{\label{THGEM close-up}Close-up of three sectors with rim sizes  of 0.1~mm, 0.15~mm, and 0.2~mm (left to right).}
\end{figure}

The promise of 3D-printing is the potentially ultra-rapid path from designing a detector in a CAD program to its production on a printer. The ambition is to complete a design or make a design change during the day, press a button to print it out overnight, and have a new detector in hand for testing the next morning. The reality at this stage of the development of the technology is quite different. As our institution does not own a PCB printer, various prototypes are printed directly by the printer company as a courtesy. A first print by a regional technical support team resulted in a board that had the front and back of the THGEM shorted together due to conductive ink extending into the holes. Another print attempt produced a version with insulating solder mask on top the front and back metal electrodes. Ultimately, the working prototype had to be printed at the company headquarters by an expert team that was able to adjust the print parameters so that the demanding requirements on the print precision could be met. The resulting final product is shown in Figs.~\ref{THGEM light} and \ref{THGEM close-up}.

\begin{figure}
\centering
\includegraphics[height=7cm]{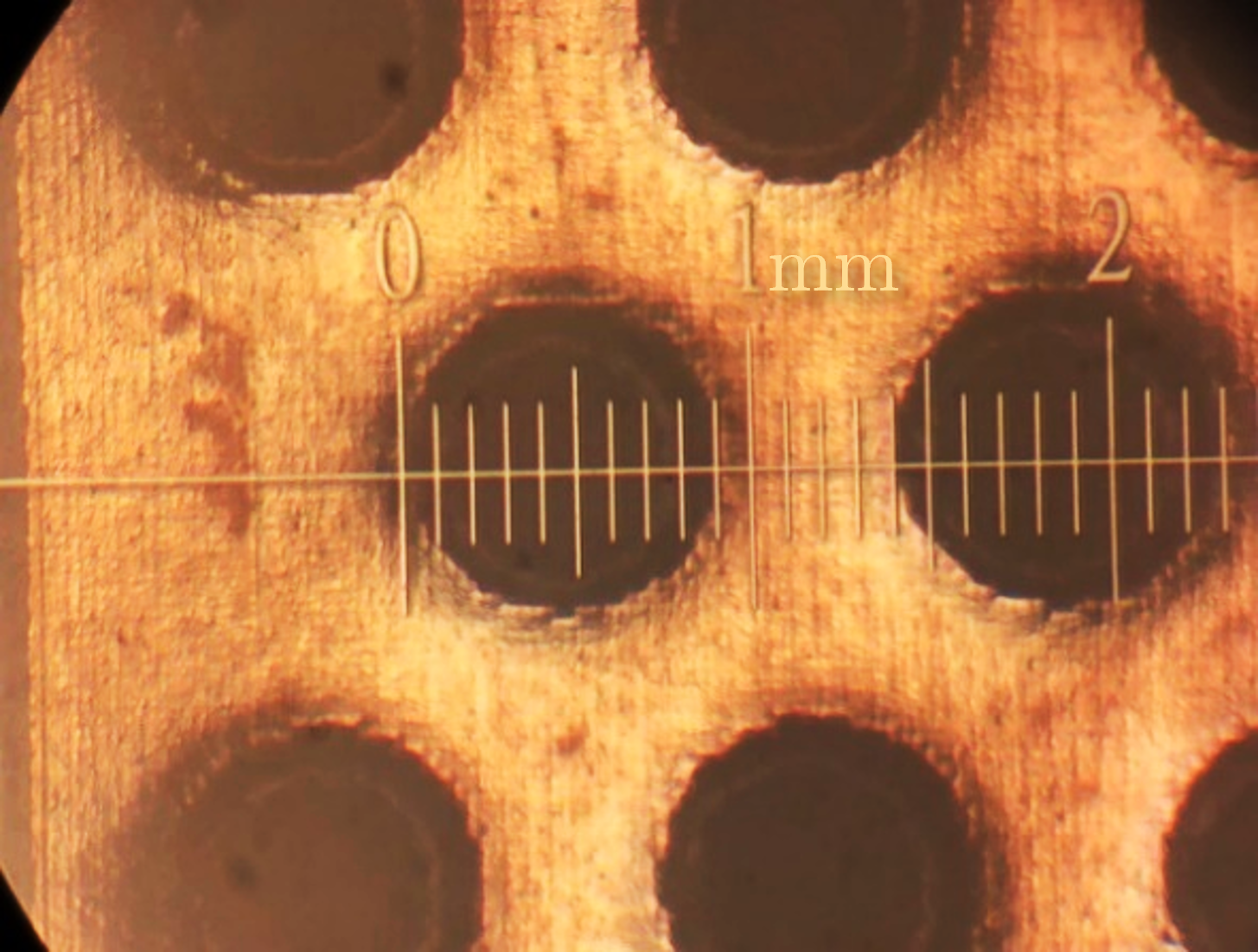}\vspace{1mm}
\includegraphics[height=7cm]{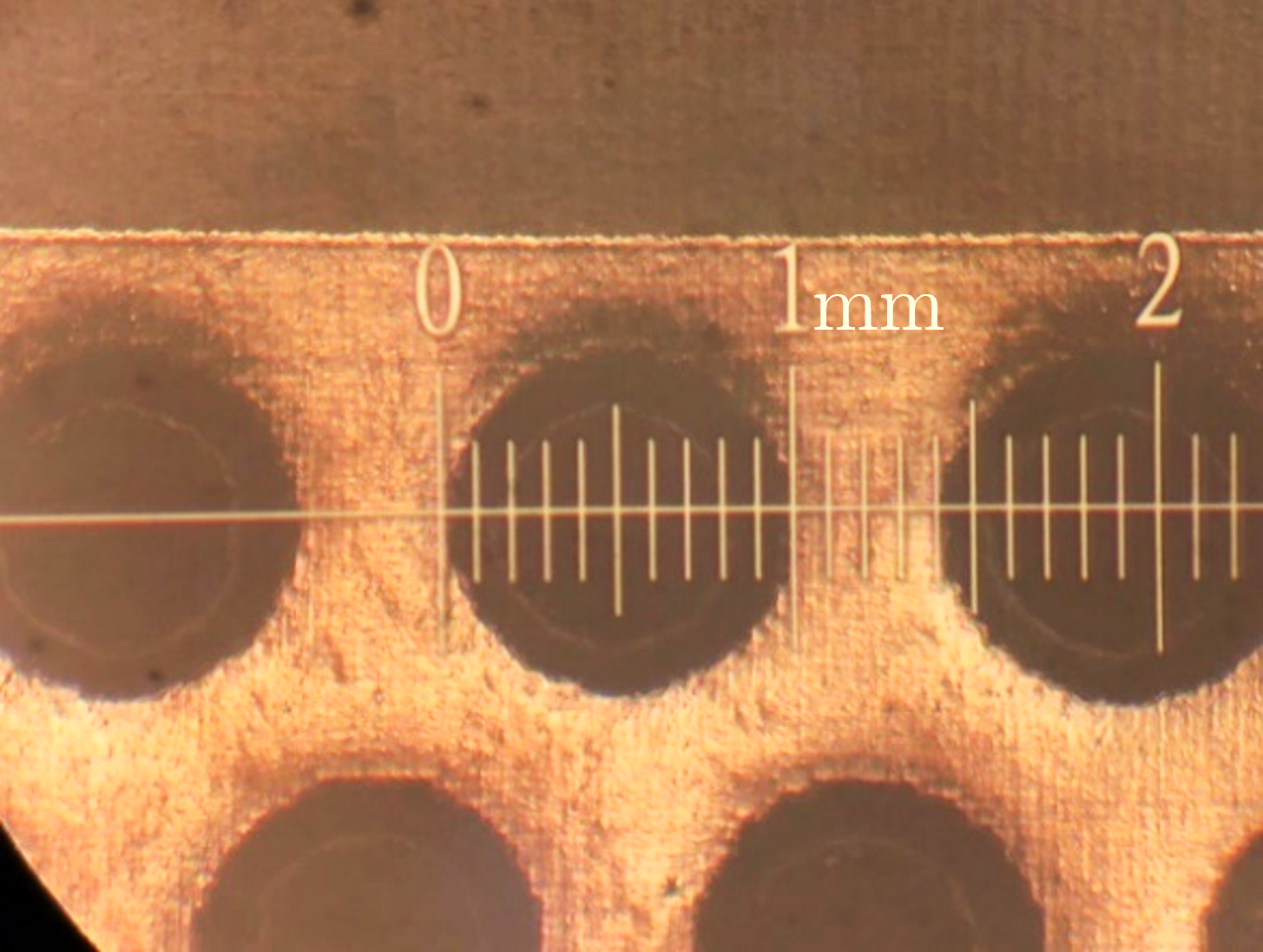}\vspace{1mm}
\includegraphics[height=7cm]{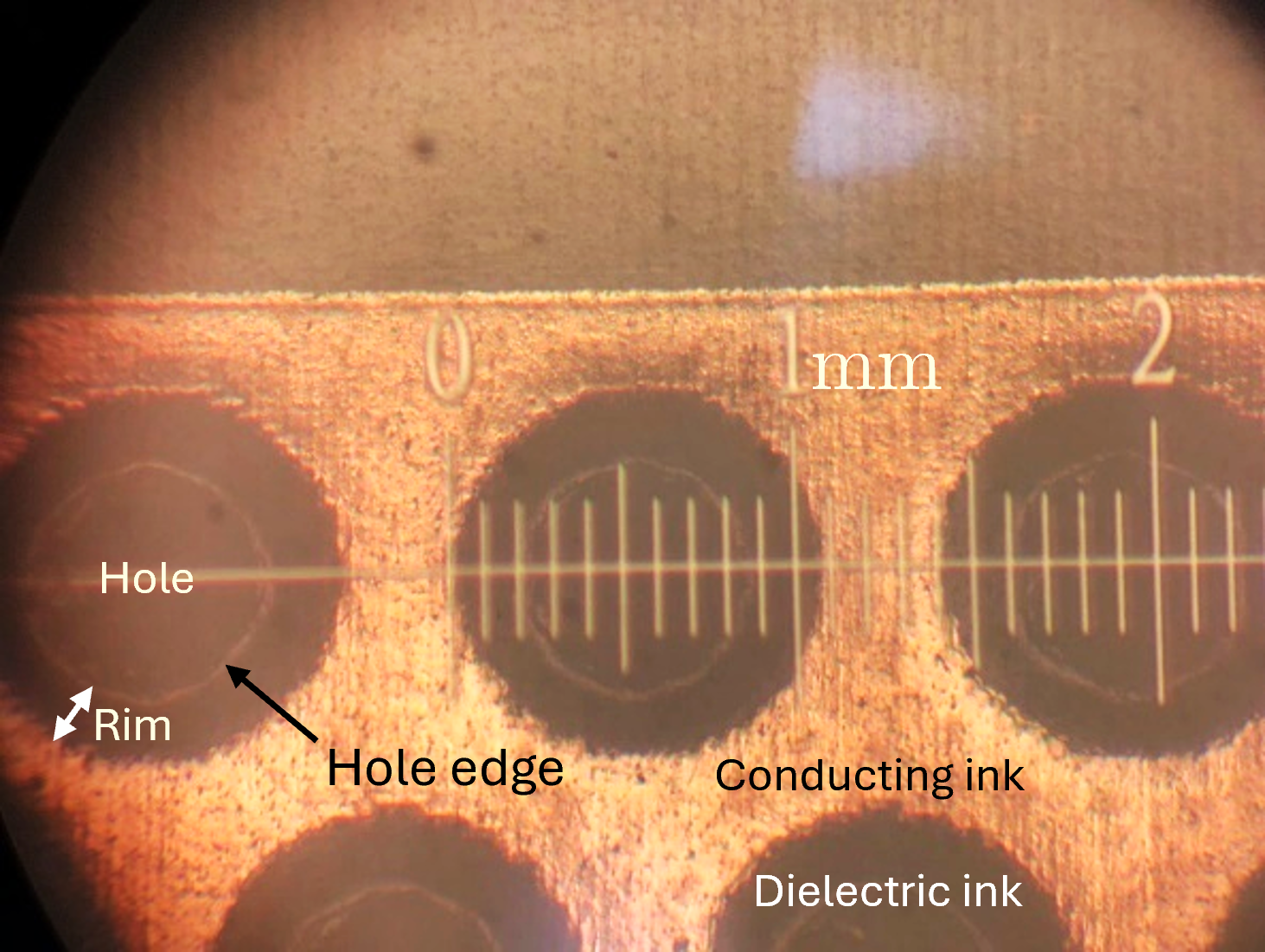}
\caption{\label{0.18 mm rims}Optical microscopy of THGEM holes. Top: Holes designed to have no rims, but with actual 0.1~mm rim annulus. Center: Holes designed to have 0.1 mm rim annulus, but with actual 0.15~mm rim annulus. Bottom: Holes designed to have 0.18 mm rim annulus, but with actual 0.2~mm annulus.}
\end{figure}

\section{Quality Control with Optical Microscopy}

The hole dimensions are spot-checked with an optical microscope against a graded glass calibration slide. Hole diameters are found to be 0.7 mm as intended; however, the rim sizes are not to specification. Instead of 0~mm, 0.1~mm, and 0.18~mm rim annuli, the actual annuli are found to be 0.1~mm, 0.15~mm, and 0.2~mm, respectively, as shown in Fig.~\ref{0.18 mm rims}. This is most likely due to shrinking of the 3D-printed material after printing. Our chief objective of determining whether or not a 3D-printed thick GEM can be used by itself to produce a functioning particle detector, is not significantly hindered by this; however, the parameters of the actual board have to be taken into consideration instead of the design parameters when evaluating the performance results.

\section{Detector Assembly}

\begin{figure}[H]
\centering
\includegraphics[width=0.9\linewidth]{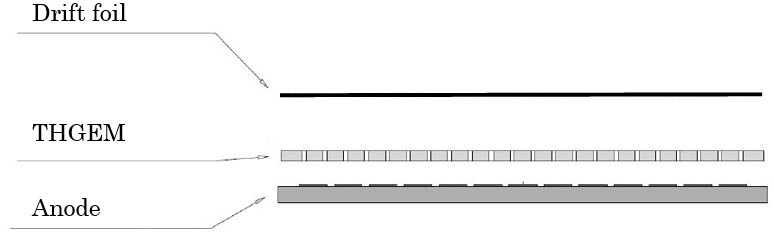}
\caption{\label{THGEM Stack graphic}A typical single-THGEM detector stack (reproduced from \cite{Alexeev:2015kda}).}
\end{figure}

\begin{figure}[H]
\centering
\includegraphics[width=0.9\linewidth]{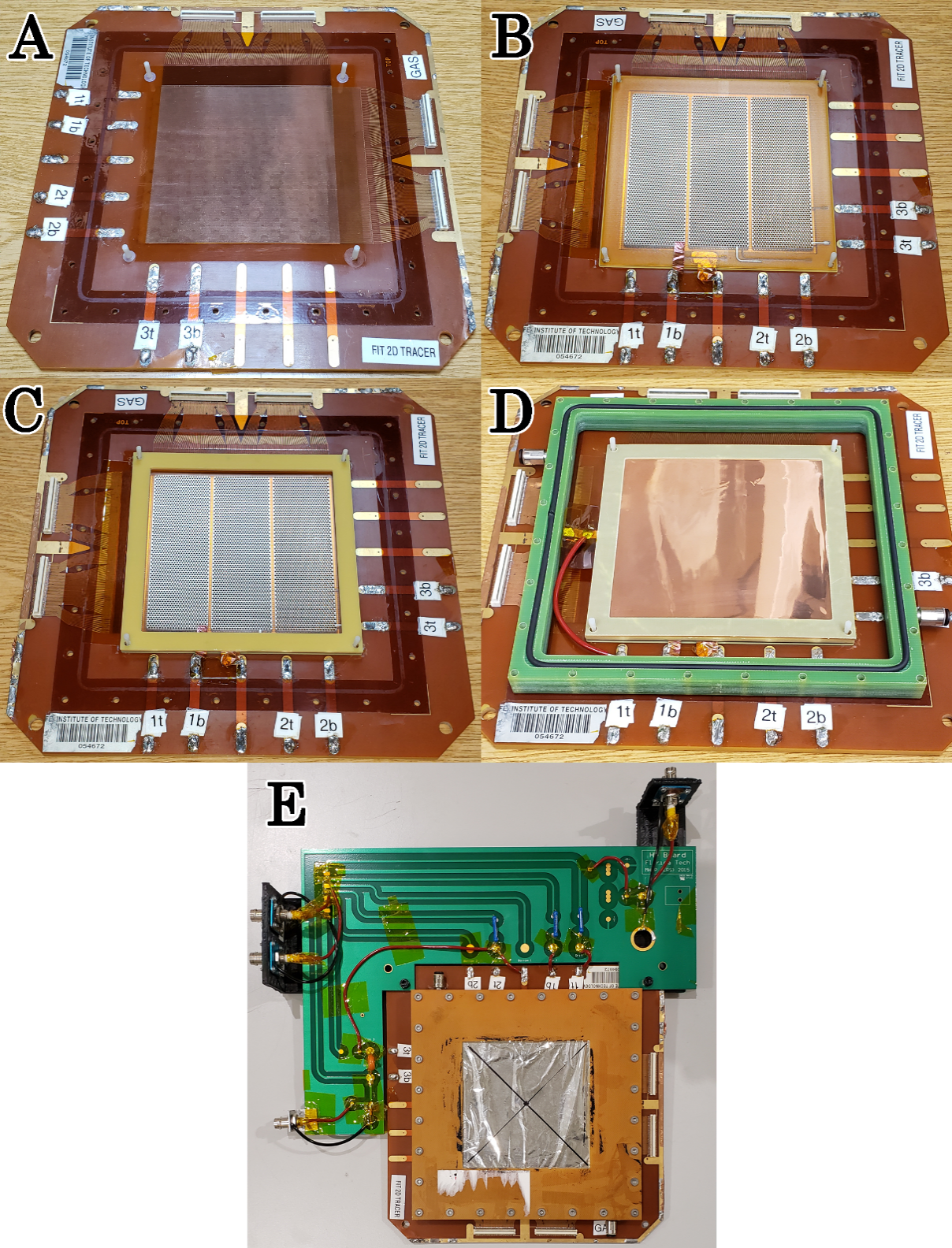}
\caption{\label{THGEM Detector Assembly} THGEM Assembly: A) X/Y-directional straight strips serve as the readout anode and deliver the signal via traces to associated Panasonic connectors outside the gas envelope. B) The THGEM board is installed atop 1~mm washers in each corner, which define the induction gap, and suspended above the readout strips. C) Three 1~mm frames are stacked above the THGEM to define the 3~mm drift gap. D) The drift cathode foil is placed atop these frames, with fixation nuts screwed onto the nylon posts, thus completing the THGEM stack. An outer frame (green) with O-rings for sealing is placed around the stack. E) Assembly of the THGEM detector is completed by placing a thin window over the active area and attaching an HV board.}
\end{figure}

The THGEM board itself is only one component needed to produce a functioning detector. A drift cathode, readout board, and gas envelope are also required. An example schematic of such an electrode stack is shown in Fig.~\ref{THGEM Stack graphic}. Taking inspiration from the short induction gap in the “3/1/2/1~mm (drift/transfer1/transfer2/induction)” configuration of the CMS GEM detectors~\cite{CMSMuonTDR}, the drift gap between the drift cathode and the top of the THGEM is set to 3~mm and the induction gap between the bottom of the THGEM and the anode readout board is set to 1~mm. The readout structure is a standard 10~cm $\times$ 10~cm x-y strip readout as originally designed and produced for the COMPASS GEMs~\cite{Altunbas:2002ds}. The main steps of the detector assembly are shown in Fig.~\ref{THGEM Detector Assembly}.

Each electrode, i.e.\ drift cathode, and top and bottom sides of the THGEM, is powered independently with HV to gain full control over each electric field.
Connection of the HV lines from the HV board to the THGEM is established using wide copper tape that is connected to the HV pads on the THGEM and soldered on both ends.

\section{Testing and Results}

\begin{figure}[H]
\centering
\includegraphics[width=0.9\linewidth]{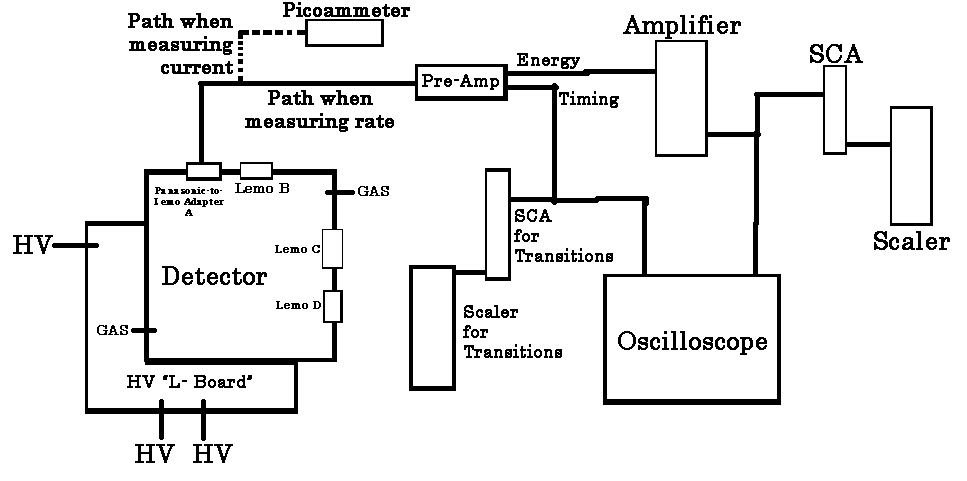}
\caption{\label{Readout electronics chain}Readout electronics chain used during testing. Note: Energy and Timing outputs from the pre-amp are just labels; the signal out of both is identical. SCA: Single Channel Analyzer used as discriminator.}
\end{figure}

The schematic shown in Fig.~\ref{Readout electronics chain} depicts the readout electronics chain used during testing of the detector. Three individual HV channels power the three electrodes of the THGEM. Sets of 128 readout strips are connected via PCB traces to a Panasonic connector, with two sets of strips in the X direction and two sets in the Y direction, labeled as sectors A, B, C, D. For gain measurements, the strips in each set are ganged together externally using a small Panasonic-to-Lemo  adapter board. 

The ganged signal is routed to an ORTEC pre-amplifier in the case of rate measurements or to a picoammeter for anode current measurements. The pre-amplifier has two identical output channels (labeled ``timing'' and ``energy''). One channel is split and connected to a single-channel analyzer (SCA), whose discriminated output is connected to a scaler for counting large positive-to-negative transitions from potential sparks, and to an oscilloscope for pulse monitoring. The signal from the other pre-amp channel is sent to an amplifier, whose output is split to another SCA that leads to a scaler, for counting signal pulses, and to an oscilloscope for signal pulse monitoring. For the purposes of these tests, only one set of 128 ganged strips at a time is read out; the others are terminated with 50 Ohm terminators to ground them so that the electric field in the induction gap is properly maintained everywhere.

The testing of the THGEM Detector comprises four main tests: A test to determine if the board contains any shorts, a basic test of THGEM functionality by establishing presence of signal pulses, a test to determine the gain, and a test of the long-term gain behavior.

The first test is done for all three sectors of the THGEM board to verify that they are free of shorts. An impedance of at least 100 G$\Omega$ is found with a Gigaohmmeter when a 500 Volt potential difference is held across both sides of a THGEM sector for a duration of at least 5 minutes. From this result we conclude that there are no shorts of any kind on the THGEM board.

\begin{figure}[t]
    \centering 
    \includegraphics[trim=0 0 1mm 0, clip, width=0.7\linewidth]{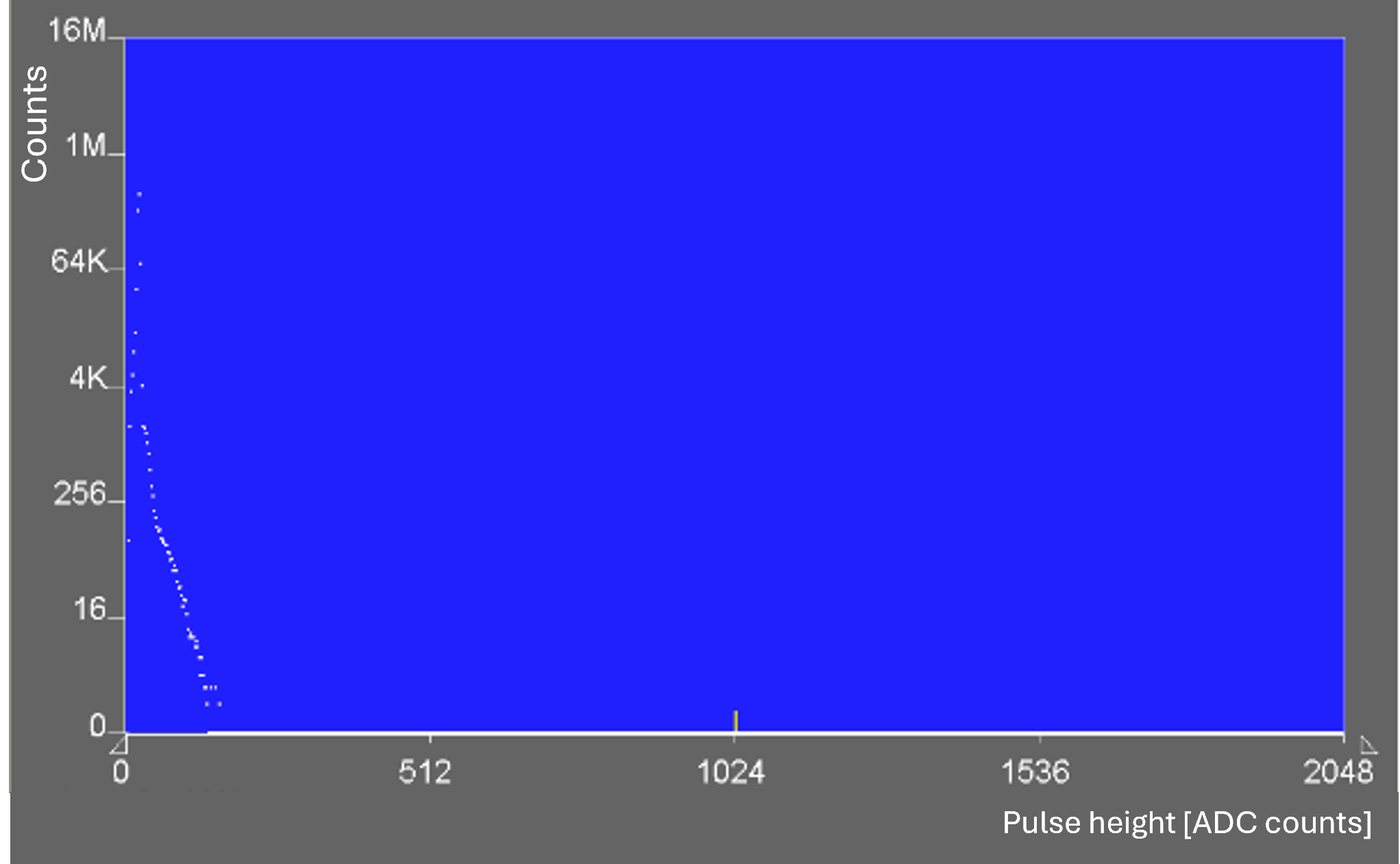}\vspace{2mm}
    \includegraphics[trim=0 0 1mm 0, clip, width=0.7\linewidth]{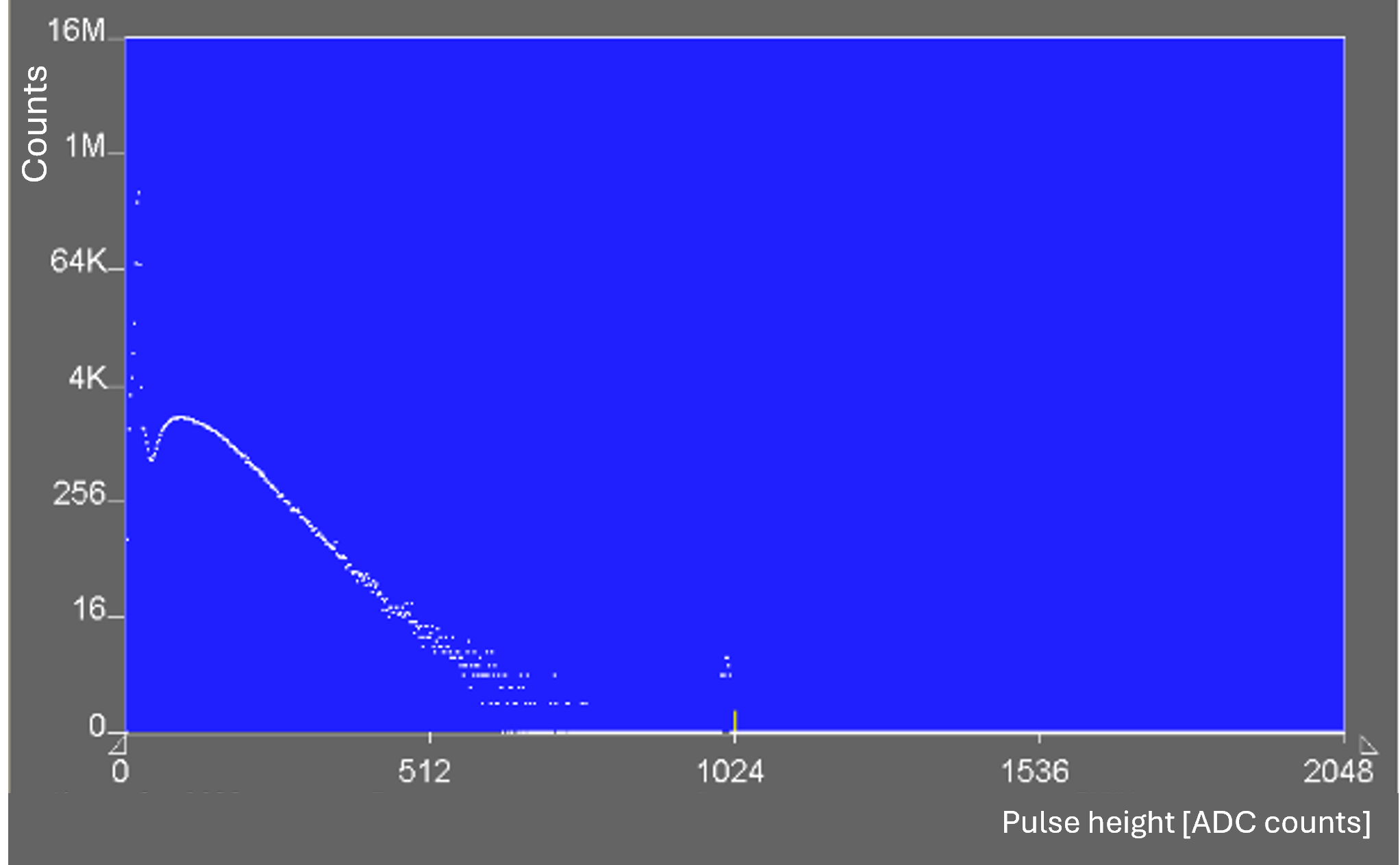}
    \caption{\label{MCA}Pulse height spectra in ADC counts of amplifier output pulses for the THGEM sector with 0.15~mm rim annuli using a multi-channel analyzer. Top: Without source. Bottom: With an $^{241}$Am source irradiating the detector.}   
\end{figure}

\subsection{Establishing basic functionality of Thick-GEM detector}

We check if the 3D-printed THGEM produces detectable signal pulses at the output of the pre-amp and amplifier with the oscilloscope when a highly-ionizing $^{241}$Am source irradiates the drift cathode through the thin detector window. The detector is flushed with an Ar/CO$_2$ 70:30 gas mixture. Drift and induction fields are both set to typical values used in MPGDs and the bias voltage across the THGEM is slowly increased to see if an amplified output signal can be detected.

The first positive observation is made for the THGEM sector with 0.1~mm rim annuli operated with a 3.8~kV/cm drift field and a 2~kV/cm induction field; under those conditions sizable pulses begin to appear on the oscilloscope with a THGEM bias voltage around 1800~V. When removing the sources from the detector, we observe that these pulses disappear. Pulse height distributions for the sector with 0.15~mm rim annuli measured with and without an $^{241}$Am alpha source using a multichannel analyzer show a clear difference; the distribution taken with source extends to significantly higher pulse heights (Fig.~\ref{MCA}). This establishes that these pulses are indeed due to the detection of particles from the sources and not due to some discharge process. Similar behavior is observed for all three THGEM sectors albeit at different values for the THGEM bias voltage. We conclude that the 3D-printed THGEM acts indeed as an active gain element.

\subsection{Gas gain measurement}
We proceed to measure the gas gain vs.\ THGEM bias voltage for each of the THGEM sectors. For these tests we use an Amptek Mini-X X-ray gun with a Cu filter and a small $^{55}$Fe source. The detector is placed inside a lead-lined box that houses the X-ray gun. An interlock system ensures that if the box is accidentally opened while the X-ray is on, the gun is powered down. The X-ray gun is placed approximately 1 cm from the gas window of the detector.

\begin{figure}[b]
\centering
\includegraphics[width=0.9\linewidth]{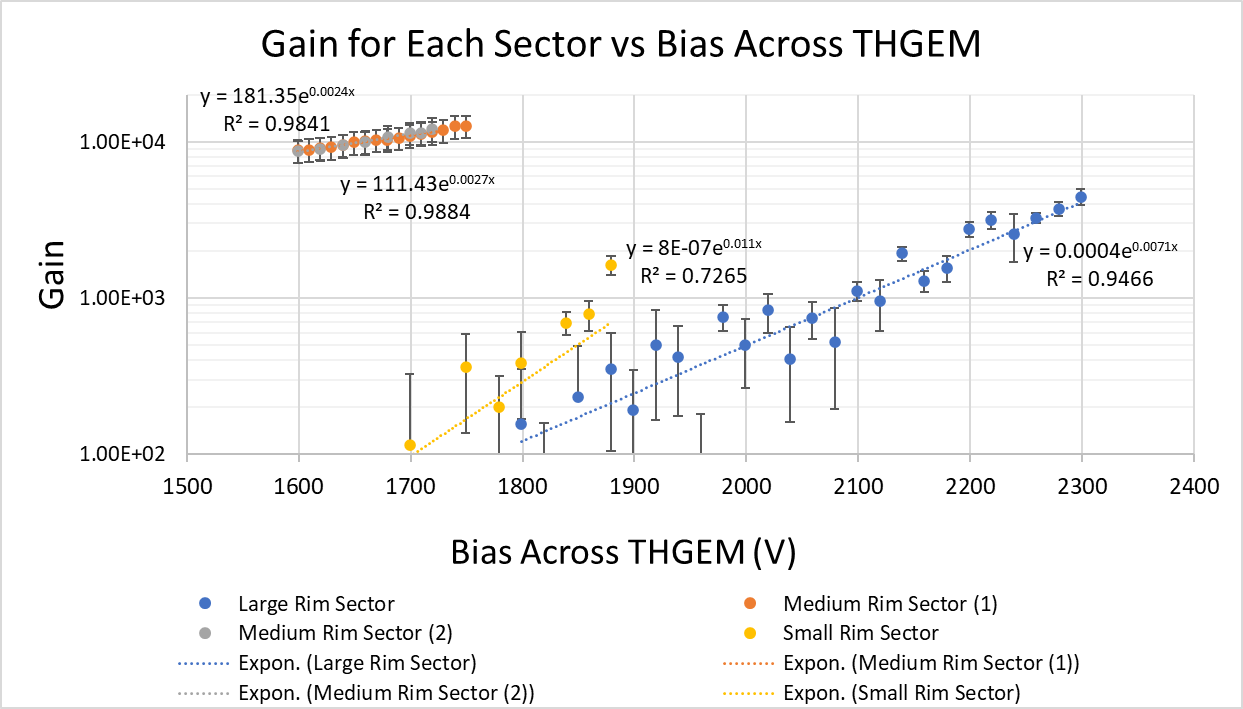}
\caption{\label{Gain curves}Gain curves from fits to an exponential for each region of the board in Ar/CO$_2$ (70:30). The small and large rim sectors are tested using an $^{55}$Fe source, while the medium rim sector is tested using an X-ray gun with Cu filter. Gain curves from two separate trials are shown for the medium THGEM sector with 0.15~mm rim annuli.} 
\end{figure} 

For the central THGEM sector with medium (0.15~mm) rim annuli, the pulse rate and anode current are measured up to a THGEM bias voltage of 1750V using 2.16~kV/cm and 1.8~kV/cm induction and drift fields, respectively, which allow stable operation. We calculate the gas gain G from these measurements using the well-known formula $G=\frac{I}{n_{prim}\cdot e\cdot R_i}$, where I is the anode current, $R_i$ is the incident rate of X-rays, $n_{prim}$ is the number of primary electron-ion pairs due to ionization, and $e$ is the elementary charge. For Ar/CO$_2$ (70:30), $n_{prim} = 298$ for 8~keV Cu fluorescence X-rays. Typically, the incident rate is determined from a plateau in such rate measurement. However, the THGEM experiences discharges before a rate plateau is reached. 
Instead, here the incident rate is measured independently with a Geiger counter that replaces the THGEM detector at the same distance to the source and is found to be 1.8~kHz. 

The gain curves for all three sectors measured at 2.16~kV/cm and 1.8~kV/cm induction and drift fields, respectively, are compiled in Fig.~\ref{Gain curves}. These exponential curves cover a THGEM bias voltage range where operation was stable and end where discharges became too frequent for reasonable operation. 
At a given THGEM bias voltage, the medium-rim THGEM sector achieves higher gas gains than the other two sectors; it is also the only sector to achieve a gain above $10^4$ with stable operation. The small-rim sector barely achieves a gain of 1000 before becoming unstable.

For the most promising THGEM sector with medium rims, at a fixed THGEM bias voltage of 1703V, we vary the induction field while keeping the drift field constant and vice versa and measure the gas gain. We observe a linear dependence of the gain on the induction field over the range $E_{\textrm{ind}} = 1.25-3.25$ kV/cm and on the drift  field over the range $E_{\textrm{drift}} = 1-2.25$ kV/cm. The gain vs.\ drift field plateaus over the range $E_{\textrm{drift}} = 2.25-3 $ kV/cm~\cite{JerryThesis}.

\subsection{Gain stability tests}

\begin{figure}[b]
\centering
\includegraphics[width=0.9\linewidth]{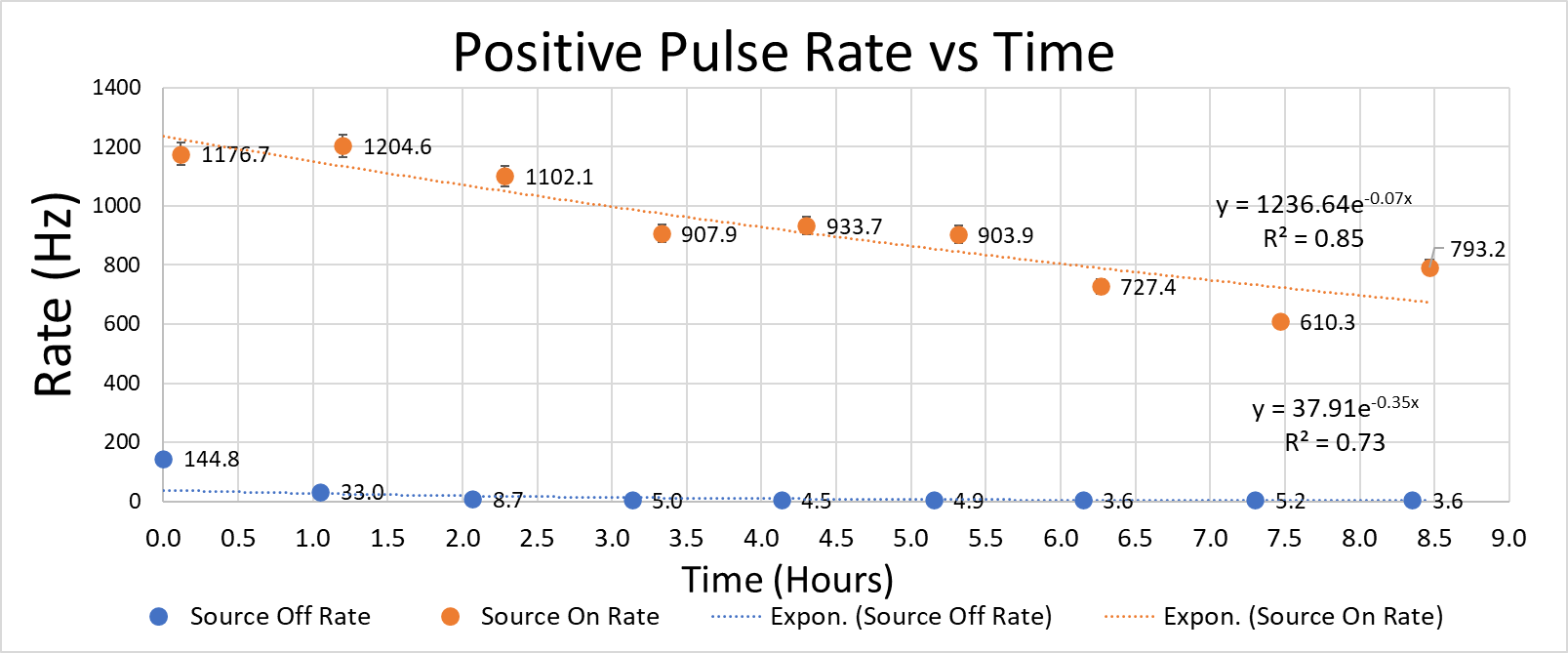}
\caption{\label{Positive pulses over time}Rate of signal pulses over time for the medium rim sector in Ar/CO$_2$ (70:30). Error bars are present for all points, but relatively small compared to axis units.}
\end{figure}

We conduct some longer-term irradiation tests to investigate the stability of the gas gain. These tests have the THGEM under $^{55}$Fe irradiation for a period of 5-48 hours. Observations are made regarding the HV stability (trips), count rate, gain, and the rate of the occurrence of large positive-to-negative “transitions” in the pre-amp signal, which are considered as due to discharges.
In these tests, the sector with medium-size rims is studied using drift and induction fields of 3~kV/cm and a bias voltage across the THGEM of 2000~V. For the first long-term test, the detector is kept under HV continuosly, but is only irradiated briefly at certain times to measure the count rates. Fig.\ref{Positive pulses over time} shows a general trend of the observed count rate decreasing by almost 50\% over a period of about eight hours.

In a second, longer test over 48 hours, the $^{55}$Fe source is irradiating the detector continuously. Count rates are measured more frequently for approximately the first two hours, then once per hour for the next six hours and not at all overnight. The following day, rates are measured every four hours until the late evening, and the last two measurements are completed the day after.

\begin{figure}[t]
\centering
\includegraphics[width=0.9\linewidth]{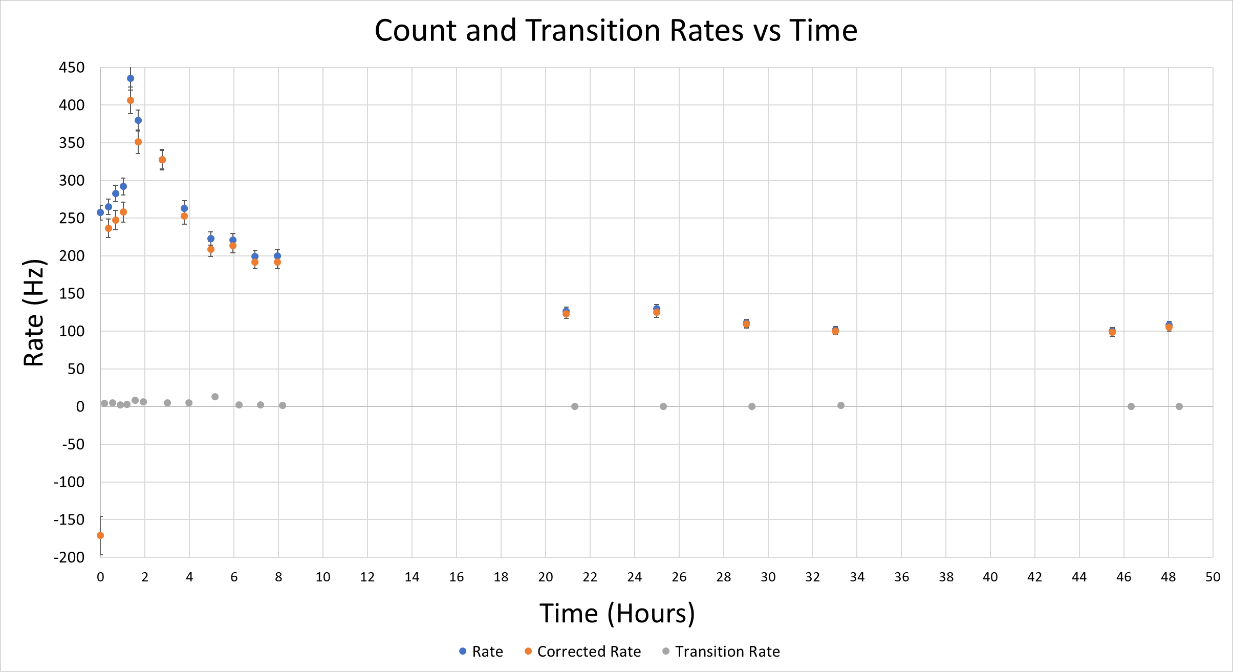}
\caption{\label{Count and Transition rates vs time 48 hour}Raw rate (blue) and rate corrected for discharges (orange) for regular signal pulses and rate of ``transition'' pulses (gray) in first test over a 48-hour period for the medium rim sector in Ar/CO$_2$ (70:30). Error bars are present for all points, but relatively small compared to axis units.}
\end{figure}

Fig.\ref{Count and Transition rates vs time 48 hour} shows that the count rate rises quickly in the first 90 minutes from initially 240~Hz to a maximum of 400~Hz before beginning to fall until reaching a plateau of around 100~Hz after approximately 24 hours. We correct the measured raw rate by subtracting the rate of irregular transition pulses, presumably due to discharges, that have overshoots large enough to exceed the discriminator threshold (Fig.~\ref{Readout electronics chain}). The rate of these discharge pulses appears to get smaller over time so that, around the time of the plateau region, the raw rate and the corrected rate are approximately the same value.

To check if this peaking and then falling of the rates is a genuine manner of behavior, or just something observed once at that time, the experiment is repeated in order to confirm. This is also done because we unfortunately do not monitor changes in air pressure during the test that can impact gain and hence count rate for a fixed discriminator threshold. 
As the primary objective is to investigate the behavior with an initial rise and subsequent fall, the duration of the second long-term test is reduced to several hours. However, unlike in the previous test, in addition to the rate measurement, this time the signal current is also monitored throughout testing so that, by utilizing a reference source-off current and rate from the end of the trial, combined with the average count rate of those taken during the study, an approximate detector gain over time can be determined and monitored over several hours.

\begin{figure}
\centering
\includegraphics[width=0.9\linewidth]{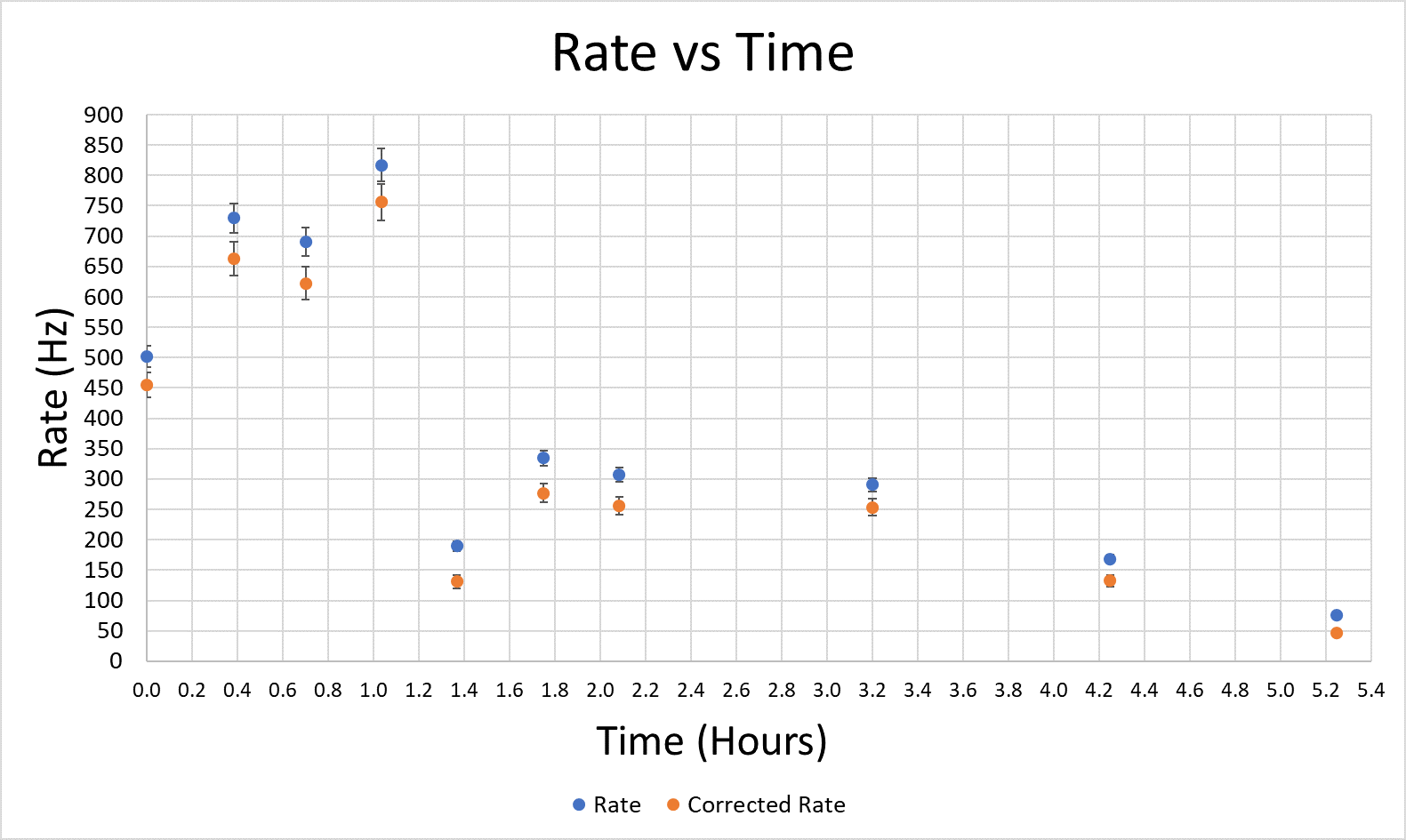}
\caption{\label{Count rate vs time} Raw rate (blue) and rate corrected for discharges (orange) for signal pulses in second long-term test for the medium rim sector in Ar/CO$_2$ (70:30). Error bars are present for all points but, in some cases, small compared to axis units.}
\end{figure}

\begin{figure}
\centering
\includegraphics[width=0.9\linewidth]{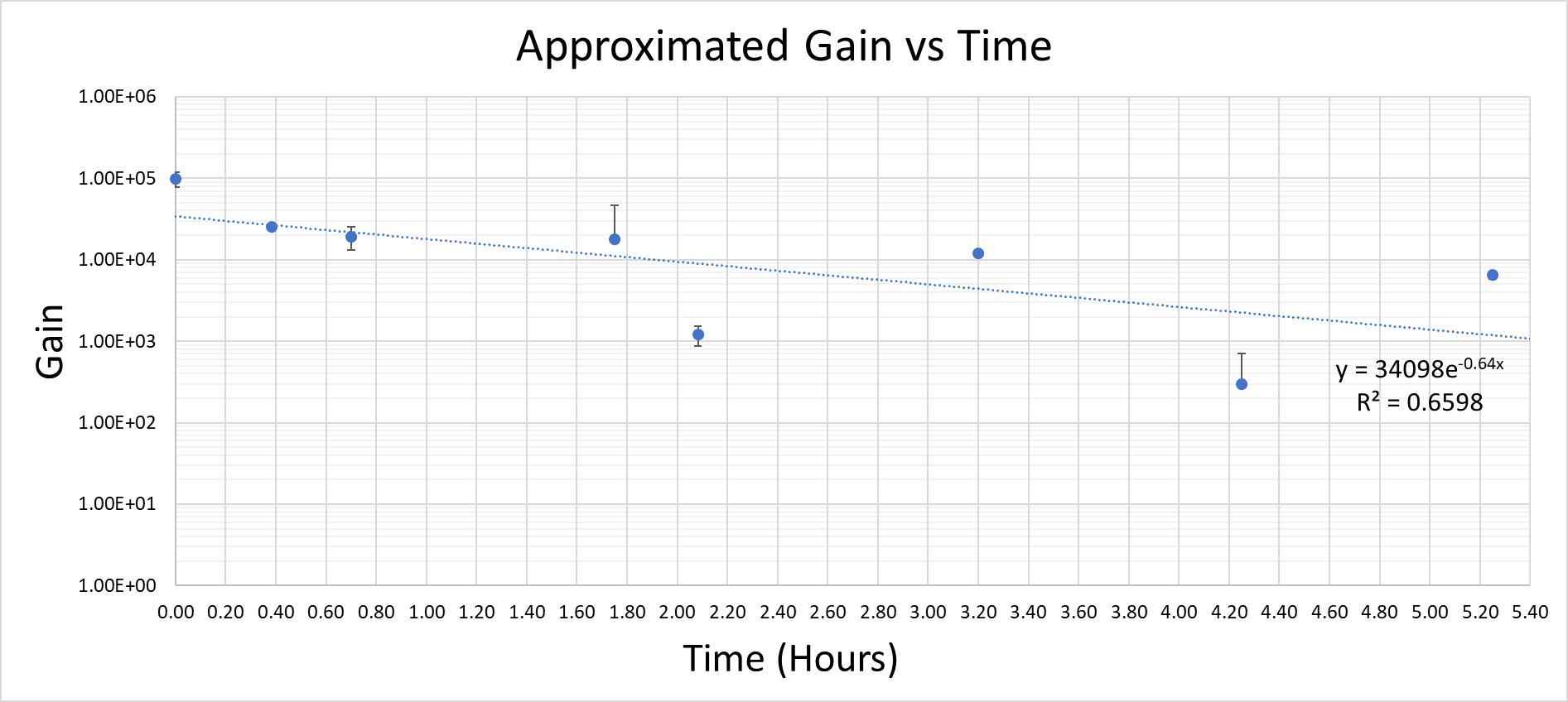}
\caption{\label{Approximated Gain vs time}Approximate gain over time during the second long-term study for the medium rim sector in Ar/CO$_2$ (70:30). Error bars are present but, in some cases, small compared to axis units. Missing lower bars are the result of not being able to cross zero on a log plot.}
\end{figure}

Fig.\ref{Count rate vs time} shows that the behavior of the signal count rate over time indeed appears to be reproducible, i.e.\ it is similar to that seen in Fig.\ref{Count and Transition rates vs time 48 hour} during the first several hours of the 48-hour study. This confirms that the rate increases during approximately the first hour under HV and irradiation and then afterwards it begins to decrease below the initial value. A corresponding approximate gain measurement for the same time period of the second test is shown in Fig.~\ref{Approximated Gain vs time}. The errors are relatively high in this measurement as the current produced with the $^{55}$Fe source is small and subject to fluctuations. The observed gain decreases quickly in the first half hour from an initially rather high value of 10$^5$ and keeps decreasing -- albeit at a smaller rate over the following five hours. This is consistent with the result from the rate measurement. We hypothesize that this behavior is due to a complex charging-up process occurring in the 3D-printed insulating THGEM material over time.

\section{Conclusions and Outlook}
We conclude that we have achieved proof-of-principle. A THGEM can be 3D-printed and achieve large gas gain exceeding 10$^4$. This allows it to function as the sole gain element in an MPGD. As with standard THGEMs, we find that the gain is critically dependent on the size of the hole rims. The 3D-printing process must be carefully controlled by printer experts to ensure high THGEM quality; it remains to be seen if regular 3D-printer end users will also be able to produce THGEM printouts of sufficient quality when 3D-printing technology for PCBs progresses. The strong time dependence of the gas gain is inconvenient for the stable operation of a detector, but other MPGDs behave in a similar way due to charging-up of the electrodes. In any future application some intial conditioning time has to be allocated to allow the gain to settle before operations begin. As a next R\&D step, one would have to investigate if a THGEM structure can be reliably printed over an area significantly larger than 10 cm $\times$ 10 cm. The size of the printer build area will ultimately present a limitation here. It would be interesting to see if the 3D-printed THGEM can be combined with a 3D-printed readout structure, possibly in a single printout as a combined unit. This would be the next step towards making the dream of printing out an entire detector at the press of a ``print'' button a reality.

\section{Acknowledgments}
We thank Florida Tech undergraduates John Hammond and Omar Nour for their help with the quality control of the THGEM and the assembly of the detector. We are grateful to NANODIM for making their expertise and printing samples available to us.

\bibliography{references}

@inproceedings{Hohlmann:2013hda,
    author = "Hohlmann, M.",
    title = "{Printing out Particle Detectors with 3D-Printers, a Potentially Transformational Advance for HEP Instrumentation}",
    booktitle = "{Snowmass 2013}: {Snowmass on the Mississippi}",
    eprint = "1309.0842",
    archivePrefix = "arXiv",
    primaryClass = "physics.ins-det",
    reportNumber = "SNOW13-00137",
    month = "9",
    year = "2013"
}

@article{Brunbauer:2019ubp,
    author = {Brunbauer, F. M. and Lupberger, M. and M\"uller, H. and Oliveri, E. and Pfeiffer, D. and Ropelewski, L. and Scharenberg, L. and Thuiner, P. and Van Stenis, M.},
    title = "{3D printing of gaseous radiation detectors}",
    journal = "JINST",
    volume = "14",
    number = "12",
    pages = "P12005",
    year = "2019",
    doi = "10.1088/1748-0221/14/12/P12005"
}

@article{Sauli:1997qp,
    author = "Sauli, F.",
    title = "{GEM: A new concept for electron amplification in gas detectors}",
    doi = "10.1016/S0168-9002(96)01172-2",
    journal = "Nucl. Instrum. Meth. A",
    volume = "386",
    pages = "531--534",
    year = "1997"
}

@article{Sauli:2016eeu,
    author = "Sauli, Fabio",
    title = "{The gas electron multiplier (GEM): Operating principles and applications}",
    doi = "10.1016/j.nima.2015.07.060",
    journal = "Nucl. Instrum. Meth. A",
    volume = "805",
    pages = "2--24",
    year = "2016"
}

@book{Sauli:2021,
  author    = {F. Sauli},
  title     = {Micro-Pattern Gaseous Detectors - Principles of Operation and	Applications},
  publisher = {World Scientific},
  year      = {2021},
  edition   = {$1^{\textrm{st}}$},  
  address   = {Singapore}, 
  isbn      = {978-981-12-2221-4},
  doi = {10.1142/11882}
}

@article{Breskin:2008cb,
    author = "Breskin, A. and Alon, R. and Cortesi, M. and Chechik, R. and Miyamoto, J. and Dangendorf, V. and Maia, J. and Dos Santos, J. M. F.",
    editor = "Buzulutskov, A.",
    title = "{A concise review on THGEM detectors}",
    eprint = "0807.2026",
    archivePrefix = "arXiv",
    primaryClass = "physics.ins-det",
    doi = "10.1016/j.nima.2008.08.062",
    journal = "Nucl. Instrum. Meth. A",
    volume = "598",
    pages = "107--111",
    year = "2009"
}

@article{Alexeev:2015kda,
    author = "Alexeev, M. and others",
    title = "{The gain in Thick GEM multipliers and its time-evolution}",
    doi = "10.1088/1748-0221/10/03/P03026",
    journal = "JINST",
    volume = "10",
    number = "03",
    pages = "P03026",
    year = "2015"
}

@article{Bondar:2012qz,
    author = "Bondar, A. and Buzulutskov, A. and Dolgov, A. and Grebenuk, A. and Shemyakina, E. and Sokolov, A. and Akimov, D. and Breskin, A. and Thers, D.",
    title = "{Two-phase Cryogenic Avalanche Detectors with THGEM and hybrid THGEM/GEM multipliers operated in Ar and Ar+N2}",
    eprint = "1210.0649",
    archivePrefix = "arXiv",
    primaryClass = "physics.ins-det",
    doi = "10.1088/1748-0221/8/02/P02008",
    journal = "JINST",
    volume = "8",
    pages = "P02008",
    year = "2013"
}

@misc{nanodim,
  title        = "{DragonFly 3D-printer}",
  author       = "{NANODIMENSION, Inc.}",
  howpublished = "\url{https://www.nano-di.com/dragonfly-iv}"
}

@misc{altium,
  title        = "{Altium Design Software}",
  author       = "{Altium, Inc.}",
  howpublished = "\url{https://www.altium.com/}"
}

@techreport{CMSMuonTDR,
  title       = "{The Phase-2 Upgrade of the CMS Muon Detectors}",
  author      = "{CMS Collaboration}",
  institution = "CERN",
  number      = "CERN-LHCC-2017-012, CMS-TDR-016",
  howpublished = "\url{https://cds.cern.ch/record/2283189}",
  year        = 2017
}

@article{Altunbas:2002ds,
    author = "Altunbas, C. and others",
    title = "{Construction, test and commissioning of the triple-GEM tracking detector for COMPASS}",
    reportNumber = "CERN-EP-2002-008",
    doi = "10.1016/S0168-9002(02)00910-5",
    journal = "Nucl. Instrum. Meth. A",
    volume = "490",
    pages = "177--203",
    year = "2002"
}

@mastersthesis{JerryThesis,
  author  = "{Jerry L. Collins II}",
  title   = "{Design, Assembly, and Testing of a Small 3D-printed Thick-GEM}",
  school  = "Florida Institute of Technology",
  year    = 2021,
  address = "Melbourne, FL",
  month   = Dec
}

\end{document}